\documentclass[pre,twocolumn,superscriptaddress,nofootinbib]{revtex4-2}
\usepackage{graphicx,amssymb,amsfonts,amsmath,chemarr,color,commath,braket,hyperref}

\begin{document}

\title{Restoring information in aged gene regulatory networks by single knock-ins}

\author{Ryan LeFebre}
\affiliation{Department of Physics and Astronomy, University of Pittsburgh, Pittsburgh, Pennsylvania 15260, USA}

\author{Fabrisia Ambrosio}
\affiliation{Department of Physical Medicine and Rehabilitation, Harvard Medical School, Boston, MA, USA}
\affiliation{Discovery Center for Musculoskeletal Recovery, Schoen Adams Research Institute at Spaulding, Boston, MA, USA}

\author{Andrew Mugler}
\email{andrew.mugler@pitt.edu}
\affiliation{Department of Physics and Astronomy, University of Pittsburgh, Pittsburgh, Pennsylvania 15260, USA}

\begin{abstract}
A hallmark of aging is loss of information in gene regulatory networks. These networks are tightly connected, raising the question of whether information could be restored by perturbing single genes. We develop a simple theoretical framework for information transmission in gene regulatory networks that describes the information gained or lost when a gene is “knocked in” (exogenously expressed). Applying the framework to gene expression data from muscle cells in young and old mice, we find that single knock-ins can restore network information by up to 10\%. Our work advances the study of information flow in networks and identifies potential gene targets for rejuvenation.
\end{abstract}

\maketitle

\section{Introduction}
\begin{figure}[t]
\includegraphics[width=.7\columnwidth]{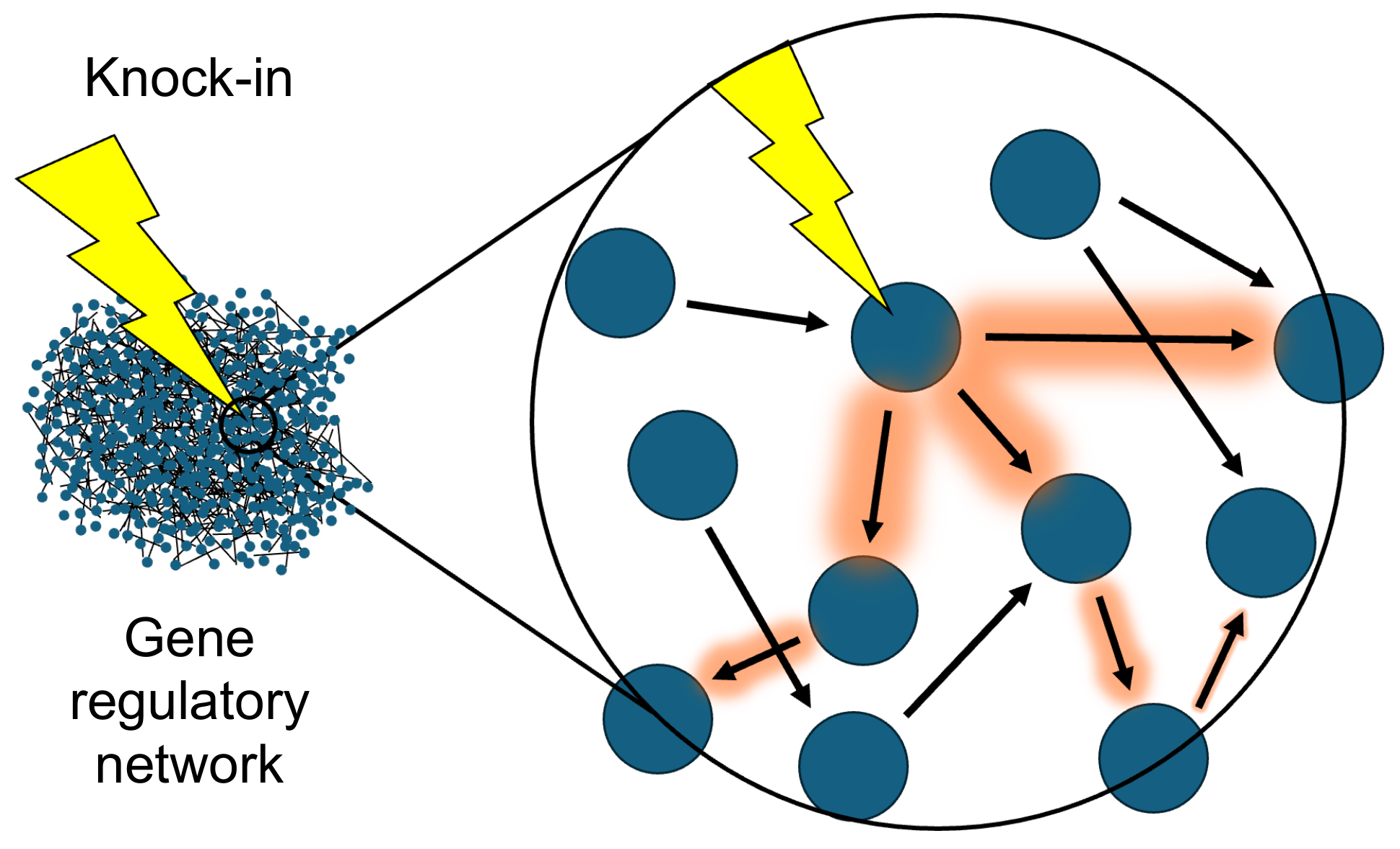}
\caption{Schematic of a gene regulatory network (blue) in which one gene is perturbed (yellow) by being ``knocked in,'' or exogenously expressed. The knock-in affects the gene's regulatory targets, and their targets, and so on (orange), suggesting that one perturbation could significantly affect information transmission across the whole network.}
\label{cartoon}
\end{figure}

Aging is a complex process characterized by hallmarks at the cellular level \cite{lopez2013hallmarks, lopez2023hallmarks}. Several of these hallmarks involve disruption of gene expression. Gene expression is a tightly coordinated and integrative process, with the products of many genes regulating the expression of other genes. These regulations form a network that encodes proper cell function (Fig.\ \ref{cartoon}, blue). The information processed by gene regulatory networks can be quantified using metrics from information theory \cite{tkavcik2008information, tostevin2009mutual, mugler2009spectral, mugler2010information, tkavcik2011information, cheong2011information, selimkhanov2014accurate, tkavcik2025information}. By these metrics, recent analyses have demonstrated that aging is associated with a loss of information within cells' gene regulatory networks \cite{southworth2009aging, clemens2021biphasic, leote2024loss, wang2025personalized, sivakumar2025novel}.

The high connectivity of this network raises the question of how much information could be restored by perturbing the expression of a single gene \cite{emison2025decay}. Specifically, one could imagine that if a gene were perturbed (Fig.\ \ref{cartoon}, yellow), the genes that it regulates would also be perturbed, and so on across the network (Fig.\ \ref{cartoon}, orange). Therefore, depending on what information is lost with age, and which genes are perturbed, the perturbation of single genes may have a significant impact on restoring information across the entire network. Perturbing the expression of single genes in the cells of living organisms is experimentally feasible, for example through gene editing, RNA interference, or epigenetic modifications.

Comprehensive data are publicly available on the regulatory links and gene expression levels in young and old organisms. In particular, databases of the gene regulatory interactions in mouse cells have been compiled from a combination of high-throughput experiments and literature curation \cite{han2018trrust, kang2022rnainter, li2025regnetwork}. Gene expression data from both young and old mice are available from single-cell RNA sequencing experiments \cite{kimmel2019murine, tabula2020single, krishnarajah2022single, lagger2023scdiffcom, xu2025developing}. A pressing open problem is to develop theoretical frameworks that can account for these data and use them to make predictions for future experiments.

Here we develop a simple theoretical framework for information transmission in gene regulatory networks. Its simplicity allows us to infer the parameters for all regulatory links, and thus the information across the entire network, from gene expression data, without fitting. We apply the framework to gene expression and regulation data from the Tabula Muris Senis \cite{tabula2020single} and TRRUST \cite{han2018trrust} databases, respectively, and find that both expression levels and information decrease with age in mouse muscle cells. We then use the framework to predict how much of the lost information is restored by ``knocking in'' (exogenously expressing) each gene. We find that single knock-ins can restore up to 10\% of the lost information, and we identify the most restorative genes. Our results establish a predictive framework for information transmission in gene networks and identify potential targets for rejuvenation interventions.

\section{Results}

\subsection{Gene expression levels decrease with age}

We obtain gene expression data from the Tabula Muris Senis (TMS) database \cite{tabula2020single}. The advantage of this database is that it archives SMART-seq2 single cell RNA-seq datasets, enabling the detection of lowly expressed genes with higher sensitivity than droplet-based methods. We focus on limb muscle cells from young mice (3 months) and old mice (24 months). The data are gene transcript numbers per cell from single-cell RNA sequencing experiments, log-transformed and normalized according to the TMS protocol \cite{tabula2020single}. We obtain the mouse gene regulatory network from the Transcriptional Regulatory Relationships Unravelled by Sentence-based Text-mining (TRRUST) version 2 database. TRRUST v2 is the preferred database because it provides a high-confidence, literature-curated mouse transcriptional regulatory network. The overlap between the two databases used consists of 2,362 genes with 6,253 regulatory links. These data and all code for this work are freely available \cite{code}.

\begin{figure}
\includegraphics[width=\columnwidth]{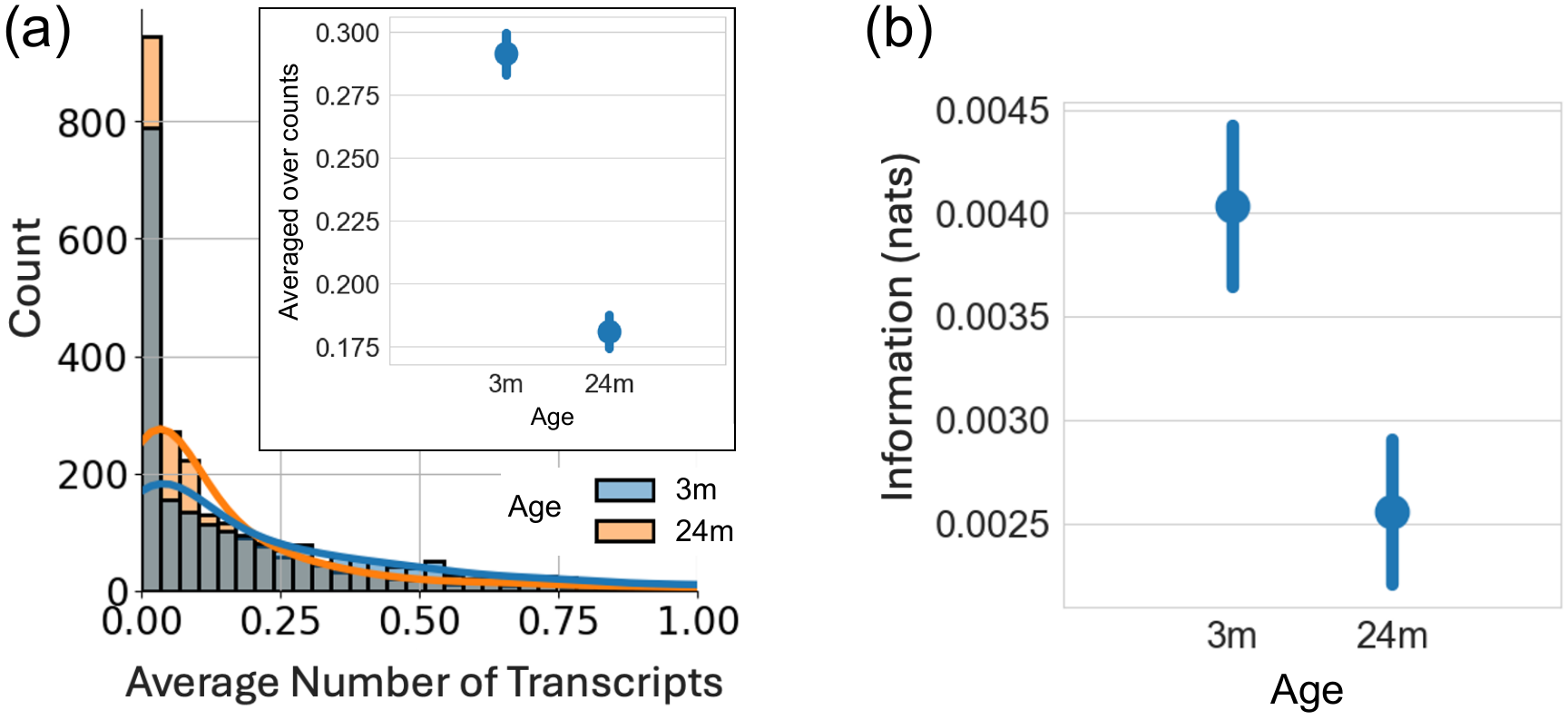}
\caption{Gene expression and information transmission decrease with age. (a) Histograms of processed transcript numbers from the TMS database \cite{tabula2020single} for limb muscle cells of young (3 months) and old (24 months) mice. Transcript numbers are averaged over 1,102 (young) or 1,232 cells (old) and binned across 2,362 genes. Curves: Gaussian kernel density estimates to guide the eye. Inset: Average and standard error of each histogram. (b) Mutual information (Eq.\ \ref{mi}), averaged over all 6,253 TF-TG pairs, for young and old mice. Error bars: standard error.}
\label{data}
\end{figure}

Figure \ref{data}(a) shows distributions of the processed transcript numbers, averaged over cells, across the 2,362 genes. We see that transcript numbers shift to lower values with age. Indeed, the inset shows the average and standard error of these distributions, and we see a substantial decrease from 3 to 24 months. Thus, we see from these data that gene expression decreases with age on average, consistent with previous findings \cite{sivakumar2025novel}.

\subsection{Information transmission decreases with age}

Single-cell RNA sequencing data is sparse \cite{lahnemann2020eleven}, meaning that many genes, in many cells, have no detected transcripts. To simplify our analysis, we adopt the previously used approach \cite{qiu2020embracing, bouland2023consequences, sarra2025maximum, sivakumar2025novel} of binarizing the data: setting all nonzero transcript numbers to one. This approach has two advantages. First, it naturally incorporates the detection threshold of the experiment into the analysis: genes that are weakly expressed but not detected fall into our ``off'' category (0), and genes that are expressed strongly enough to be detected fall into our ``on'' category (1). Second, it makes subsequent results agnostic to the processing choices made by the TMS consortium: because their processing never takes a transcript number from zero to nonzero or vice versa, binarizing the raw data and binarizing the processed data give equivalent results.

The binarized data can then be described by a probability distribution for each pair consisting of a regulating gene (or transcription factor, TF) and a regulated gene (or target gene, TG). Specifically, the joint distribution $p_{ij}$ gives the four probabilities (1) $p_{00}$ that both the TF and the TG off, (2) $p_{10}$ that the TF is on and the TG is off, (3) $p_{01}$ that the TF is off and the TG is on, and (4) $p_{11}$ that both the TF and the TG are on.
For each gene pair, these probabilities are computed directly from the data as the proportions of cells falling into the four categories.

Given a joint probability distribution for any two random variables, the information transmitted between them is given by the mutual information \cite{shannon1948mathematical, cover1999elements}, defined
\begin{equation}
\label{mi}
I = \sum_{ij} p_{ij}\log\frac{p_{ij}}{q_ir_j},
\end{equation}
where $q_i = \sum_jp_{ij}$ and $r_j = \sum_ip_{ij}$ are the marginal distributions. Independent variables, which satisfy $p_{ij}=q_ir_j$, have zero information by Eq.\ \ref{mi}, as expected. The units of $I$ depend on the base of the log: $\log_2$ gives bits; $\ln$ gives nats. Two maximally correlated binary variables, for example a fair coin ($q_0 = q_1 = 1/2$) that perfectly predicts a second coin ($p_{00} = p_{11} = 1/2$), transmit one bit of information. We use natural log here, for which this maximum case gives $I = \ln 2\approx 0.69$ nats.

Figure \ref{data}(b) shows that the information $I$, averaged over all 6,253 TF-TG pairs, decreases with age. The average $I$ values are small (hundredths of nats), as many TF-TG pairs express little to no information in limb muscle cells. Nevertheless, the decrease in information with age is larger than the standard errors. The observation that gene regulatory information decreases with age is consistent with previous findings \cite {sivakumar2025novel, emison2025decay}.

\subsection{Theoretical framework predicts the effects of knock-ins}

\begin{figure}
\includegraphics[width=\columnwidth]{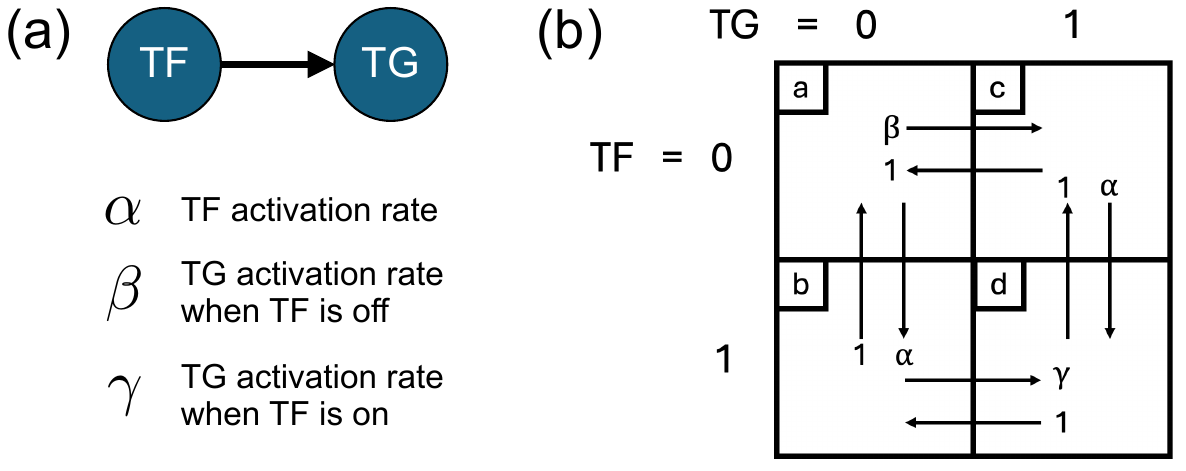}
\caption{Model of binary gene regulation. (a) A transcription factor (TF) regulates a target gene (TG). (b) Rates determine transitions between the four possible states.}
\label{model}
\end{figure}

If both gene expression and information transmission go down with age, can expressing particular genes bring information transmission back up? To address this question, we define a simple model of binary gene regulation, shown in Fig.\ \ref{model}. A TF switches from its off state to its on state at a rate $\alpha$. Meanwhile, a TG switches from its off state to its on state at a rate that depends on the TF state: $\beta$ if the TF is off, and $\gamma$ if the TF is on. If $\beta<\gamma$, then the TF activates the TG; if $\beta>\gamma$, then the TF represses the TG. All rates are scaled by the common rate at which both genes switch to their off state, which we set to one.

This model has two key features, which we describe below. First, the parameters can be inferred from the data without fitting. Second, the model admits a simple way to implement the knock-in of a gene, i.e., exogenous expression separate from the regulation.

To infer the parameters from the data, we solve the model in steady state, as follows. For brevity, we define
\begin{equation}
\label{abcd}
p_{00} \equiv a, \quad
p_{10} \equiv b, \quad
p_{01} \equiv c, \quad
p_{11} \equiv d.
\end{equation}
The dynamics of the model are given by a set of rate equations (also called a master equation) that include a term for each possible transition, illustrated in Fig.\ \ref{model}(b):
\begin{align}
\dot a &= b + c - (\alpha+\beta)a, \\
\dot b &= \alpha a + d - (1+\gamma)b, \\ 
\dot c &= \beta a + d - (1+\alpha)c, \\ 
\dot d &= \gamma b + \alpha c - 2d.
\end{align}
Setting the time derivatives to zero and solving for the probabilities, we obtain the steady state
\begin{align}
\label{a}
a &= (2 + \alpha + \gamma)/z, \\
b &= [\alpha (2 + \alpha + \beta)]/z, \\ 
c &= [\alpha \gamma + \beta(2 + \gamma)]/z, \\ 
\label{d}
d &= [\alpha\beta + \alpha\gamma(1 + \alpha + \beta)]/z,
\end{align}
where
\begin{equation}
z \equiv (1 + \alpha)[2 + \alpha + 2 \beta + \gamma(1 + \alpha + \beta)]
\end{equation}
is a normalization factor. Eqs.\ \ref{a}-\ref{d} satisfy $a+b+c+d=1$, as required by probability conservation. We invert them to solve for the parameters $\alpha$, $\beta$, and $\gamma$ in terms of the probabilities,
\begin{align}
\label{alpha}
\alpha &= \frac{b + d}{a + c}, \\
\beta &= \frac{c(a+b+c)-ad}{a(a+c)}, \\
\label{gamma}
\gamma &= \frac{d(2a+c)-bc}{b(a+c)}.
\end{align}
Now it is clear that, given $a\equiv p_{00}$, $b\equiv p_{10}$, $c\equiv p_{01}$, and $d\equiv p_{11}$ from the data, the parameters can be calculated directly by Eqs.\ \ref{alpha}-\ref{gamma}, without fitting.

Inspecting Eqs.\ \ref{alpha}-\ref{gamma}, two possible issues can arise. The first is that the denominators could be zero. None of the 6,253 TF-TG pairs has $a=0$ (i.e., in at least one cell, both TF and TG are off, such that $a\equiv p_{00}>0$). Therefore, the denominators of $\alpha$ and $\beta$ are never zero. But some pairs have $b=0$, potentially making the denominator of $\gamma$ zero. In most of these pairs, $d=0$ also. In this case, setting $d=0$ in $\gamma$ allows $b$ to drop out, leaving $\gamma=c/(a+c)$, which is finite. The remaining pairs, with $b=0$ and $d>0$, make $\gamma$ infinite, but this case is rare, occurring in only 44 of 6,253 pairs ($0.7\%$). We exclude these pairs from the analysis.

The second issue is that the numerators of $\beta$ and $\gamma$ could be negative, which does not make sense for a rate. In these cases, we set the rate to zero. We have checked that this modification is small in each case using the Kullback-Leibler divergence, a measure of the difference between two probability distributions: the divergence between $p_{ij}$ from the data and $p_{ij}$ calculated with the rate(s) set to zero, is much smaller than the average divergence between all pairs of $p_{ij}$ from the data.

The second feature, the knock-in of a gene, is implemented as follows. If the gene is a TF, we increase $\alpha$ according to
\begin{equation}
\label{tfk}
{\rm TF:}\quad \alpha \to \alpha + k.
\end{equation}
Because $\alpha$ is the endogenous expression rate, $k$ is the exogenous expression rate from the knock-in. If the gene is a TG, we increase both $\beta$ and $\gamma$ according to
\begin{equation}
\label{tgk}
{\rm TG:}\quad \beta \to \beta + k, \quad \gamma \to \gamma + k.
\end{equation}
Again, $\beta$ and $\gamma$ are the endogenous expression rates that depend on the TF state, and $k$ is the exogenous expression rate that is independent of the TF state.

By changing the rate(s), a knock-in changes the steady-state probabilities in the model according to Eqs.\ \ref{a}-\ref{d}. In particular, a knock-in increases the probability of the corresponding gene to be in its on state, defined
\begin{align}
\label{ponTF}
p^{\rm TF}_{\rm on} &= p_{10} + p_{11} = b+d, \\
\label{ponTG}
p^{\rm TG}_{\rm on} &= p_{01} + p_{11} = c+d.
\end{align}
It also changes the mutual information $I$ for the pair via Eq.\ \ref{mi}. Thus, both $p_{\rm on}$ and $I$ are functions of the knock-in strength $k$.

In summary, to predict the effect of a knock-in on the information transmitted between a TF-TG pair, we (1) calculate the rates from the experimental probabilities in old mice using Eqs.\ \ref{alpha}-\ref{gamma}, (2) implement the knock-in using Eq.\ \ref{tfk} or \ref{tgk}, (3) calculate the new probabilities using Eqs.\ \ref{a}-\ref{d}, and (4) calculate the new on-probability using Eq.\ \ref{ponTF} or \ref{ponTG} and the new mutual information using Eq.\ \ref{mi}. Repeating this procedure for many values of the knock-in strength allows us to predict how the information depends on the on-probability across all knock-in strengths.

\subsection{Single knock-ins restore information up to 10\%}
\label{one}

The above framework allows us to investigate the effect of a single knock-in on a single TF-TG pair. However, we seek the effect of a single knock-in on the entire network. We achieve this extension in two steps. First, in this section, we consider all pairs in which the knocked-in gene participates. We call these pairs ``distance 1.'' Second, in the next section, we consider all pairs connected to those pairs (distance 2), and then to those pairs (distance 3), and so on throughout the network.

\begin{figure}
\includegraphics[width=\columnwidth]{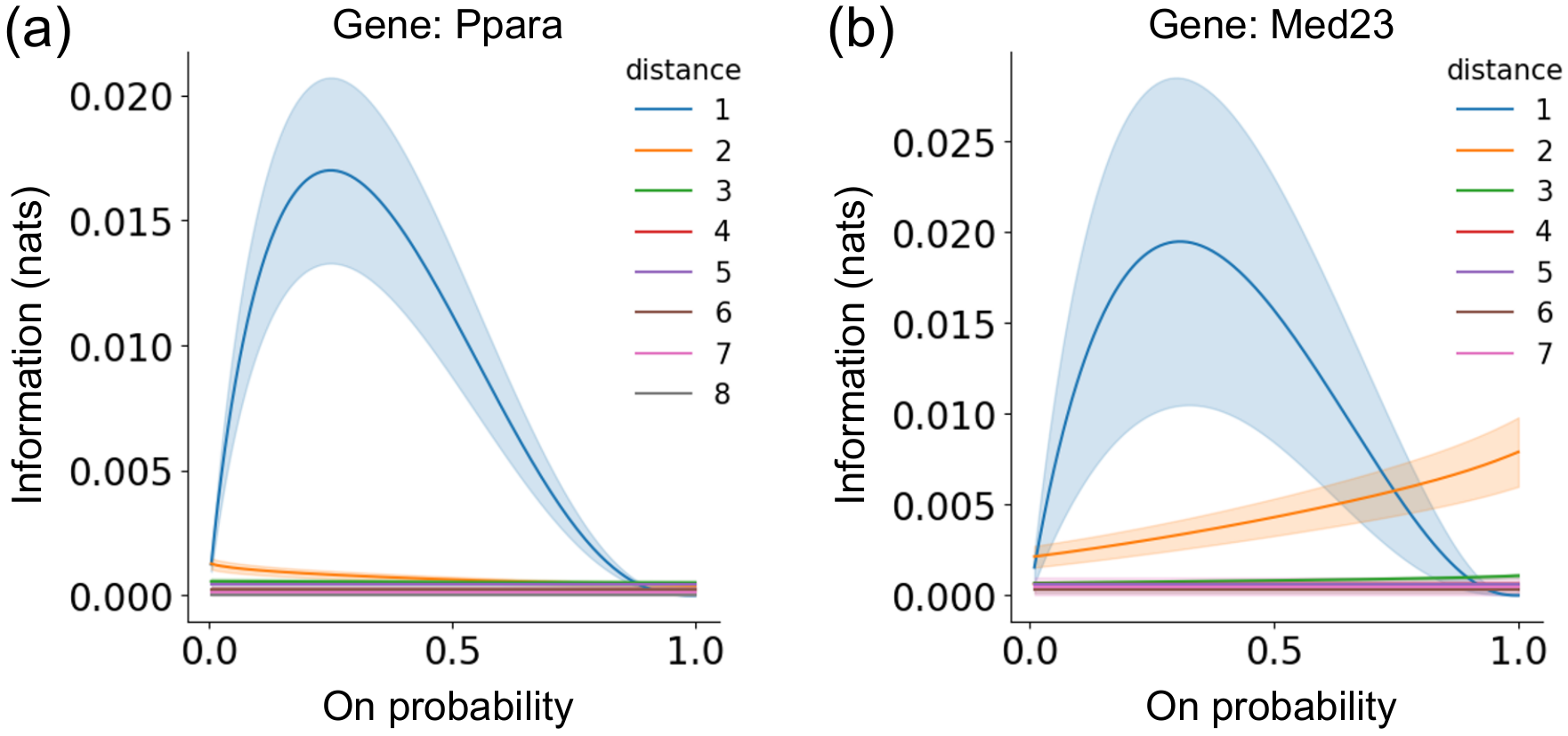}
\caption{Information vs.\ on-probability for two example knocked-in genes. Each curve is the average and standard error of $I$ (Eq.\ \ref{mi}) vs.\ $p_{\rm on}$ (Eqs.\ \ref{ponTF}, \ref{ponTG}) for all regulatory pairs a given distance from the gene in the regulatory network. The blue curves are discussed in Sec.\ \ref{one}, and all curves are discussed in Sec.\ \ref{prop}.}
\label{ip}
\end{figure}

The blue curves in Fig.\ \ref{ip} show examples for two of the most restorative genes under the protocols of this and the next section: (a) Ppara and (b) Med23. Ppara has a total of 51 identified genes that either regulate it or that it regulates. Thus, as Ppara is knocked in and its on-probability increases, there are 51 pairs whose information changes. The blue curve in Fig.\ \ref{ip}(a) shows the average and standard error of these curves. Med23 regulates 3 genes and has no regulators, and the blue curve in Fig.\ \ref{ip}(b) shows the average and standard error of its three curves.

Both blue curves in Fig.\ \ref{ip} (i) start at a low on-probability, (ii) increase in information initially, and (iii) go to zero information as the on-probability approaches one. Only the third feature is guaranteed for all genes. The reason is that a fully knocked-in gene is always on, regardless of the state of the other gene in its pair. Therefore, no information is transmitted from the latter to the former, and $I=0$. This feature is a consequence of the binary nature of the model, but it is also only true for pairs at distance 1, as we will see shortly.

\begin{figure}
\includegraphics[width=\columnwidth]{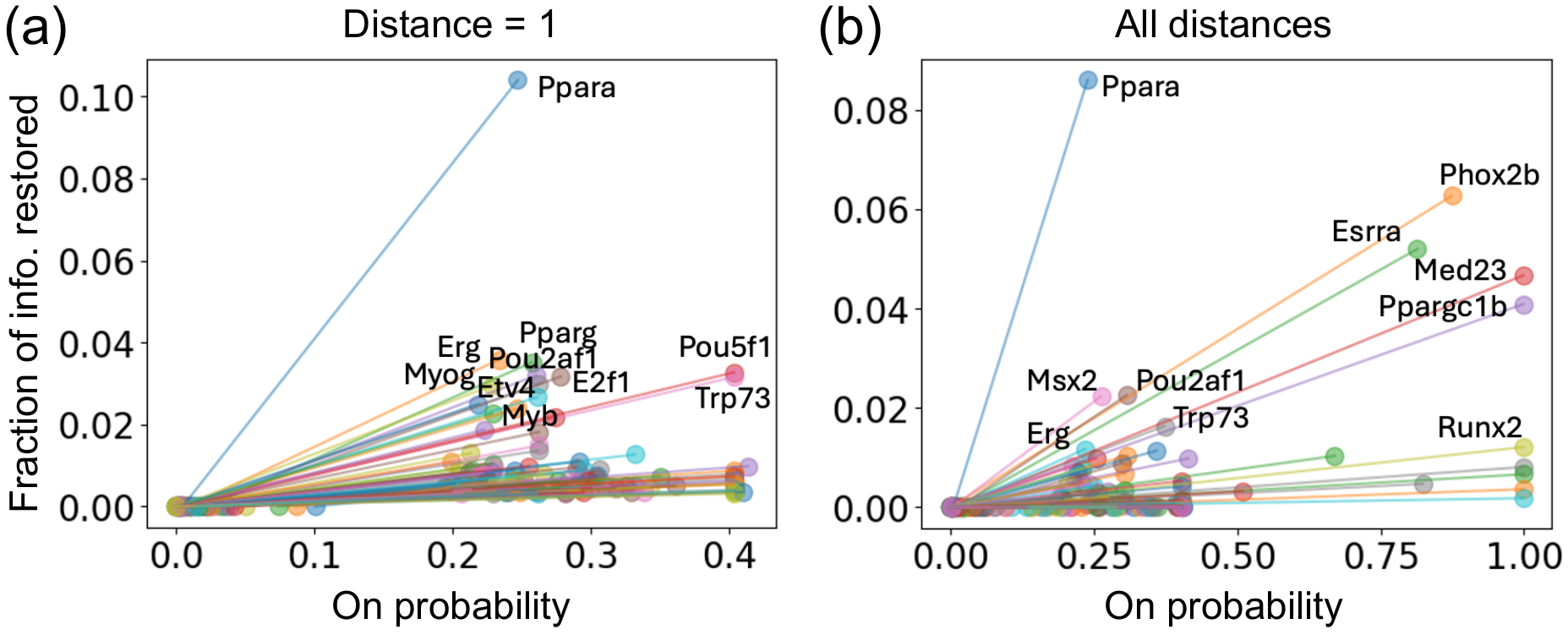}
\caption{Maximal fraction of information restored by knock-in, and the on-probability at which it occurs, for the 100 most restorative genes. (a) Considering only regulations in which the knocked-in gene participates (distance 1). (b) Propagating the knock-in to the whole network (all distances). In b, 10 repeats are performed (see Sec.\ \ref{prop}); standard error of resulting information values is smaller than the data points.}
\label{lollipop}
\end{figure}

The blue curve in Fig.\ \ref{ip}(a) indicates that the average amount of information restored is maximized at about $0.02$ nats per affected TF-TG pair, when Ppara is expressed with a probability of about 25\%. This is about 10 times the amount of information lost per pair in the entire network due to age: about $0.0015$ nats, seen in Fig.\ \ref{data}(b). However, the Ppara knock-in restores information to only about 1\% of pairs under the current protocol (51 pairs out of 6,253). The other pairs are unaffected, and so the average amount of restored information per pair in the entire network is about $10\times1\%=10\%$ of that lost. This value of 10\%, and the 25\% on-probability, are shown in Fig.\ \ref{lollipop}(a) (blue line and point labeled Ppara).

The blue curve in Fig.\ \ref{ip}(b) indicates a similar maximum amount of restored information per affected pair when knocking in Med23, but recall that Med23 only directly affects three pairs. Thus, across the entire network, the average amount of information per pair restored by knocking in Med23 is only a tiny fraction of that lost due to age. Its point is one of many at the bottom of Fig.\ \ref{lollipop}(a).

Altogether, Fig.\ \ref{lollipop}(a) shows that, considering only pairs at distance 1, (i) a knock-in can restore information by up to roughly 10\%, and (ii) optimal on-probabilities are 40\% or less. We will see in the next section that only the first feature remains true when we allow the effects of a knock-in to propagate to all pairwise distances. Indeed, a number of genes will be predicted to restore substantial information with nearly 100\% on-probability.

\subsection{Information propagation identifies the most restorative genes}
\label{prop}

\begin{figure}
\includegraphics[width=\columnwidth]{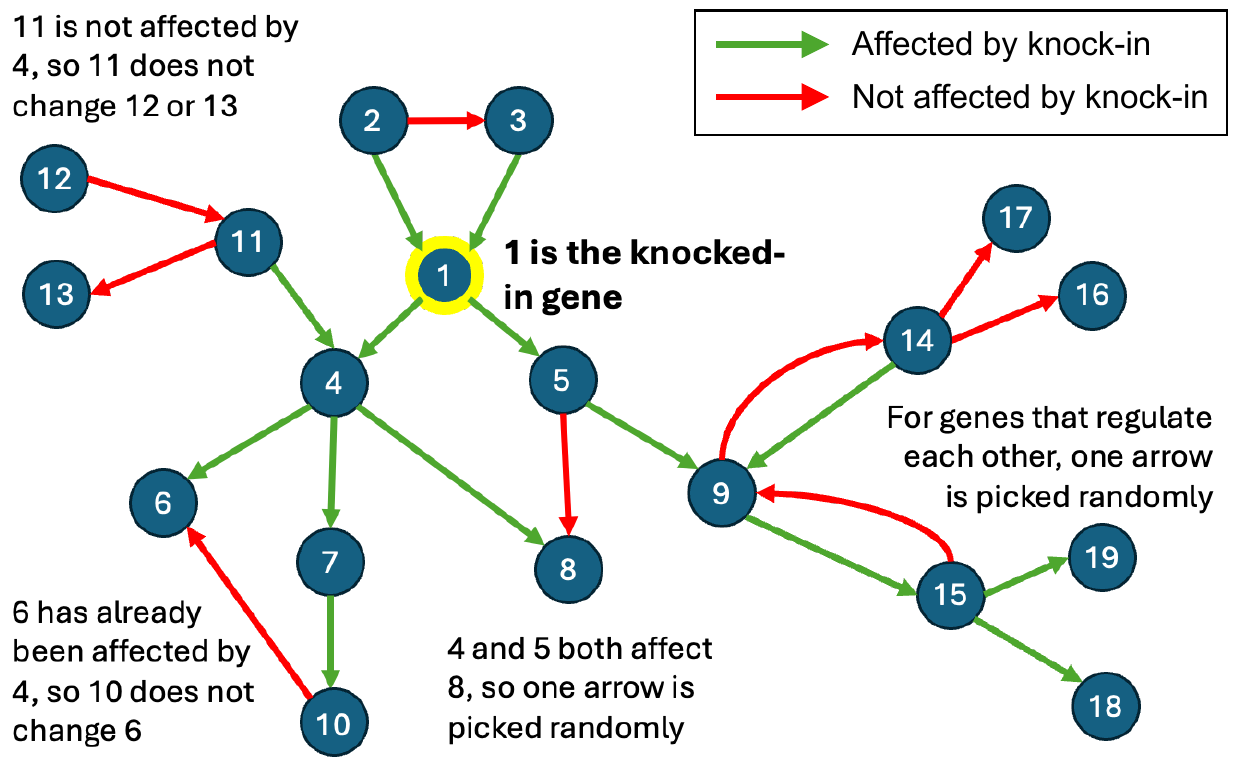}
\caption{Illustration of the algorithm used to propagate the effect of a knock-in across the network, with example steps (see description in Sec.\ \ref{prop}).}
\label{arrows}
\end{figure}

We now propagate the effect of a knock-in to all pairwise distances. Our approach is illustrated in Fig.\ \ref{arrows}. Specifically, we take the following steps. First, we compute the information for any pairs in which the knocked-in gene is the target ($2\to1$ and $3\to1$ in Fig.\ \ref{arrows}). Because the knock-in does not change the expression of the regulators in these pairs, these paths are complete. Next, we compute the information for any pairs in which the knocked-in gene is the regulator ($1\to4$ and $1\to5$ in Fig.\ \ref{arrows}). In these pairs, the expression of the regulated gene is affected: it is as if that gene is effectively, albeit more weakly, knocked in (or out). Therefore, we repeat the above two steps with each of those regulated genes ($4$ and $5$ in Fig.\ \ref{arrows}). For example, for gene 4, the first step affects the pair $11\to4$, and the second step affects the pairs $4\to6$, $4\to7$, and $4\to8$. Altogether, this repetition defines a recursive process, which we continue along each path until reaching a gene with no targets.

Two exceptions to the above algorithm arise for which our method must account: feedback and multi-input regulation. Feedback arises when a gene is encountered whose expression has been modified by the algorithm already. In this case the regulation leading to that gene is left unmodified. An example is seen in Fig.\ \ref{arrows}: gene 10 does not modify gene 6 (red arrow) because gene 4 already modified gene 6 (green arrow) in an earlier layer of the recursion (equivalently, at a smaller distance). If the feedback is direct (two genes regulate each other), then one regulation is chosen at random, and the algorithm is applied accordingly. Examples are seen in Fig.\ \ref{arrows}: one of two regulations is chosen (green arrow) for each of the pairs $9\leftrightarrow14$ and $9\leftrightarrow15$. Self-feedback is not admitted by our model, since there is no gradation in a gene's off state to influence its on rate, and vice versa.

Multi-input regulation arises when more than one gene regulates a target. In this case we choose one regulation at random, and the algorithm is applied accordingly. An example is seen in Fig.\ \ref{arrows}: genes 4 and 5 regulate gene 8; the first regulation is chosen at random (green arrow). Thus, both direct feedback and multi-input regulation involve random choice, and therefore we repeat the algorithm multiple times and average the resulting information values.

The results of this process are shown in Fig.\ \ref{ip} for the two example knocked-in genes, (a) Ppara and (b) Med23, where we separate the results based on the distance of each pair from the knocked-in gene. That is, in each case, the orange curve is the average and standard error of $I$ vs.\ $p_{\rm on}$ for all pairs a distance 2 away from the gene, the green curve for distance 3, and so on. We see that the curves flatten as the distance increases, which makes sense because the impact of the knock-in on gene expression is dampened with each successive step in the path.

We also see in Fig.\ \ref{ip}(b) that the optimal on-probability can be 100\% for distances of 2 or more (see orange curve). This makes sense because even though knocking in a gene to 100\% makes $I=0$ for any pair in which that gene is involved (blue curve), it changes the expression of that gene's targets less drastically. This smaller change may correspond to maximal information in those targets' pairs (orange curve). In fact, for Med23, there are many more pairs at distance 2 than the three pairs at distance 1, and so the orange curve dominates the overall effect of the knock-in. As a result, the average restored information is maximal at an on-probability of 100\%, as shown in Fig.\ \ref{lollipop}(b) (red line and point labeled Med23).

Altogether, Fig.\ \ref{lollipop}(b) shows that, when we propagate the effect of a knock-in to all distances, (i) a knock-in can restore information by up to roughly 10\%, and (ii) most of the highest-performing knock-ins have on-probabilities at or near 100\%. Comparing to Fig.\ \ref{lollipop}(a), we see that the second feature is new. Indeed, propagating to all distances more clearly reveals the identity of the genes whose knock-in is predicted to restore the most information: Ppara, Phox2b, Esrra, Med23, and Ppargc1b.

\section{Discussion}

Disruption of gene expression is a hallmark of aging, manifesting as a reduction in expression levels and a loss of regulatory information. We have developed a minimal model to investigate how the average information in a gene regulatory network changes when a gene is exogenously expressed, or knocked in. Applying our model to publicly available gene expression data from the limb muscle cells of young and old mice, we have found that single knock-ins in the aged network can restore information by up to 10\%, and we have predicted the most restorative genes [Fig.\ \ref{lollipop}(b)].

We find that the influence of a knock-in greatly diminishes after a few regulatory steps (Fig.\ \ref{ip}). Therefore, it is likely that if multiple genes were knocked in, then their effects on the network would have little overlap. Indeed, we have checked that among the ten most restorative genes, 73\% of the shortest connecting paths are distance 3 or more. This means that the information restored by multiple knock-ins should increase approximately additively, with little redundancy. This finding suggests that performing multiple knock-ins at once could restore information by significantly more than 10\%.

We also find that many genes restore the most information when knocked in to maximum expression levels, i.e., their on-probability approaches one in Fig.\ \ref{lollipop}(b). This prediction is experimentally convenient because it means that a knock-in can be designed to maximize gene expression rather than tuning it to a prescribed amount, which is often difficult to achieve experimentally and difficult to predict precisely from our minimal model.

Our analysis identifies the most restorative genes as Ppara, Phox2b, Esrra, Med23, and Ppargc1b. Three of these genes (Ppara, Esrra, and Ppargc1b) are connected with mitochondrial function, which is interesting given (i) the high mitochondrial content and energetic demands of skeletal muscle and (ii) the fact that mitochondrial dysfunction is an established hallmark of aging \cite{lopez2013hallmarks, lopez2023hallmarks}. Specifically, Ppara is associated with fatty acid oxidation and mitochondrial dysfunction \cite{cai2021deficiency}; Esrra expression decreases during skeletal muscle aging, paralleling loss of mitochondrial gene expression and oxphos capacity \cite{kan2021declined}; and Ppargc1b, together with PGC-1$\alpha$, drives mitochondrial biogenesis \cite{qian2024peroxisome}. Meanwhile, we are unaware of previous work connecting either of the other two genes (Phox2b and Med23) to a particular hallmark of aging. This fact demonstrates the potential of our approach to reveal novel targets that have not yet been identified in the aging field.

Several improvements are possible in future work. First, although the sparsity of the data naturally suggests it, the model could go beyond binary representation. This might prevent or delay the vanishing of information with maximal expression seen in the blue curves of Fig.\ \ref{ip}, but it would also sacrifice analytic tractability and thus computational efficiency. Second, the algorithm could be extended further to include feedback and multi-input regulation. This would represent a large increase in computational scope. Finally, we have focused here on how much information can be restored, but not necessarily where in the network it is restored (or lost in the first place). Future work could take a more localized approach in network space.

Modern high-throughput data provides comprehensive and quantitative information about holistic processes like organism aging. Interpreting these data requires targeted questions and principled theoretical frameworks. Here we have provided such a framework for investigating information transmission in aged gene regulatory networks and identifying the most promising targets for its restoration. We anticipate that our work will provide a blueprint for similar questions in the growing fielding of the physics of aging.

\acknowledgments
We thank Purushottam Dixit for helpful discussions. This work was supported by grant R01AG082739 from the National Institutes of Health.

%\bibstyle{unsrt}
%\bibliography{refs}

\begin{thebibliography}{34}%
\makeatletter
\providecommand \@ifxundefined [1]{%
 \@ifx{#1\undefined}
}%
\providecommand \@ifnum [1]{%
 \ifnum #1\expandafter \@firstoftwo
 \else \expandafter \@secondoftwo
 \fi
}%
\providecommand \@ifx [1]{%
 \ifx #1\expandafter \@firstoftwo
 \else \expandafter \@secondoftwo
 \fi
}%
\providecommand \natexlab [1]{#1}%
\providecommand \enquote  [1]{``#1''}%
\providecommand \bibnamefont  [1]{#1}%
\providecommand \bibfnamefont [1]{#1}%
\providecommand \citenamefont [1]{#1}%
\providecommand \href@noop [0]{\@secondoftwo}%
\providecommand \href [0]{\begingroup \@sanitize@url \@href}%
\providecommand \@href[1]{\@@startlink{#1}\@@href}%
\providecommand \@@href[1]{\endgroup#1\@@endlink}%
\providecommand \@sanitize@url [0]{\catcode `\\12\catcode `\$12\catcode
  `\&12\catcode `\#12\catcode `\^12\catcode `\_12\catcode `\%12\relax}%
\providecommand \@@startlink[1]{}%
\providecommand \@@endlink[0]{}%
\providecommand \url  [0]{\begingroup\@sanitize@url \@url }%
\providecommand \@url [1]{\endgroup\@href {#1}{\urlprefix }}%
\providecommand \urlprefix  [0]{URL }%
\providecommand \Eprint [0]{\href }%
\providecommand \doibase [0]{https://doi.org/}%
\providecommand \selectlanguage [0]{\@gobble}%
\providecommand \bibinfo  [0]{\@secondoftwo}%
\providecommand \bibfield  [0]{\@secondoftwo}%
\providecommand \translation [1]{[#1]}%
\providecommand \BibitemOpen [0]{}%
\providecommand \bibitemStop [0]{}%
\providecommand \bibitemNoStop [0]{.\EOS\space}%
\providecommand \EOS [0]{\spacefactor3000\relax}%
\providecommand \BibitemShut  [1]{\csname bibitem#1\endcsname}%
\let\auto@bib@innerbib\@empty
%</preamble>
\bibitem [{\citenamefont {L{\'o}pez-Ot{\'\i}n}\ \emph
  {et~al.}(2013)\citenamefont {L{\'o}pez-Ot{\'\i}n}, \citenamefont {Blasco},
  \citenamefont {Partridge}, \citenamefont {Serrano},\ and\ \citenamefont
  {Kroemer}}]{lopez2013hallmarks}%
  \BibitemOpen
  \bibfield  {author} {\bibinfo {author} {\bibfnamefont {C.}~\bibnamefont
  {L{\'o}pez-Ot{\'\i}n}}, \bibinfo {author} {\bibfnamefont {M.~A.}\
  \bibnamefont {Blasco}}, \bibinfo {author} {\bibfnamefont {L.}~\bibnamefont
  {Partridge}}, \bibinfo {author} {\bibfnamefont {M.}~\bibnamefont {Serrano}},\
  and\ \bibinfo {author} {\bibfnamefont {G.}~\bibnamefont {Kroemer}},\
  }\bibfield  {title} {\bibinfo {title} {The hallmarks of aging},\ }\href@noop
  {} {\bibfield  {journal} {\bibinfo  {journal} {Cell}\ }\textbf {\bibinfo
  {volume} {153}},\ \bibinfo {pages} {1194} (\bibinfo {year}
  {2013})}\BibitemShut {NoStop}%
\bibitem [{\citenamefont {L{\'o}pez-Ot{\'\i}n}\ \emph
  {et~al.}(2023)\citenamefont {L{\'o}pez-Ot{\'\i}n}, \citenamefont {Blasco},
  \citenamefont {Partridge}, \citenamefont {Serrano},\ and\ \citenamefont
  {Kroemer}}]{lopez2023hallmarks}%
  \BibitemOpen
  \bibfield  {author} {\bibinfo {author} {\bibfnamefont {C.}~\bibnamefont
  {L{\'o}pez-Ot{\'\i}n}}, \bibinfo {author} {\bibfnamefont {M.~A.}\
  \bibnamefont {Blasco}}, \bibinfo {author} {\bibfnamefont {L.}~\bibnamefont
  {Partridge}}, \bibinfo {author} {\bibfnamefont {M.}~\bibnamefont {Serrano}},\
  and\ \bibinfo {author} {\bibfnamefont {G.}~\bibnamefont {Kroemer}},\
  }\bibfield  {title} {\bibinfo {title} {Hallmarks of aging: An expanding
  universe},\ }\href@noop {} {\bibfield  {journal} {\bibinfo  {journal} {Cell}\
  }\textbf {\bibinfo {volume} {186}},\ \bibinfo {pages} {243} (\bibinfo {year}
  {2023})}\BibitemShut {NoStop}%
\bibitem [{\citenamefont {Tka{\v{c}}ik}\ \emph {et~al.}(2008)\citenamefont
  {Tka{\v{c}}ik}, \citenamefont {Callan~Jr},\ and\ \citenamefont
  {Bialek}}]{tkavcik2008information}%
  \BibitemOpen
  \bibfield  {author} {\bibinfo {author} {\bibfnamefont {G.}~\bibnamefont
  {Tka{\v{c}}ik}}, \bibinfo {author} {\bibfnamefont {C.~G.}\ \bibnamefont
  {Callan~Jr}},\ and\ \bibinfo {author} {\bibfnamefont {W.}~\bibnamefont
  {Bialek}},\ }\bibfield  {title} {\bibinfo {title} {Information flow and
  optimization in transcriptional regulation},\ }\href@noop {} {\bibfield
  {journal} {\bibinfo  {journal} {Proceedings of the National Academy of
  Sciences}\ }\textbf {\bibinfo {volume} {105}},\ \bibinfo {pages} {12265}
  (\bibinfo {year} {2008})}\BibitemShut {NoStop}%
\bibitem [{\citenamefont {Tostevin}\ and\ \citenamefont
  {Ten~Wolde}(2009)}]{tostevin2009mutual}%
  \BibitemOpen
  \bibfield  {author} {\bibinfo {author} {\bibfnamefont {F.}~\bibnamefont
  {Tostevin}}\ and\ \bibinfo {author} {\bibfnamefont {P.~R.}\ \bibnamefont
  {Ten~Wolde}},\ }\bibfield  {title} {\bibinfo {title} {Mutual information
  between input and output trajectories of biochemical networks},\ }\href@noop
  {} {\bibfield  {journal} {\bibinfo  {journal} {Physical review letters}\
  }\textbf {\bibinfo {volume} {102}},\ \bibinfo {pages} {218101} (\bibinfo
  {year} {2009})}\BibitemShut {NoStop}%
\bibitem [{\citenamefont {Mugler}\ \emph {et~al.}(2009)\citenamefont {Mugler},
  \citenamefont {Walczak},\ and\ \citenamefont {Wiggins}}]{mugler2009spectral}%
  \BibitemOpen
  \bibfield  {author} {\bibinfo {author} {\bibfnamefont {A.}~\bibnamefont
  {Mugler}}, \bibinfo {author} {\bibfnamefont {A.~M.}\ \bibnamefont
  {Walczak}},\ and\ \bibinfo {author} {\bibfnamefont {C.~H.}\ \bibnamefont
  {Wiggins}},\ }\bibfield  {title} {\bibinfo {title} {Spectral solutions to
  stochastic models of gene expression with bursts and regulation},\
  }\href@noop {} {\bibfield  {journal} {\bibinfo  {journal} {Physical Review
  E—Statistical, Nonlinear, and Soft Matter Physics}\ }\textbf {\bibinfo
  {volume} {80}},\ \bibinfo {pages} {041921} (\bibinfo {year}
  {2009})}\BibitemShut {NoStop}%
\bibitem [{\citenamefont {Mugler}\ \emph {et~al.}(2010)\citenamefont {Mugler},
  \citenamefont {Walczak},\ and\ \citenamefont
  {Wiggins}}]{mugler2010information}%
  \BibitemOpen
  \bibfield  {author} {\bibinfo {author} {\bibfnamefont {A.}~\bibnamefont
  {Mugler}}, \bibinfo {author} {\bibfnamefont {A.~M.}\ \bibnamefont
  {Walczak}},\ and\ \bibinfo {author} {\bibfnamefont {C.~H.}\ \bibnamefont
  {Wiggins}},\ }\bibfield  {title} {\bibinfo {title} {Information-optimal
  transcriptional response to oscillatory driving},\ }\href@noop {} {\bibfield
  {journal} {\bibinfo  {journal} {Physical review letters}\ }\textbf {\bibinfo
  {volume} {105}},\ \bibinfo {pages} {058101} (\bibinfo {year}
  {2010})}\BibitemShut {NoStop}%
\bibitem [{\citenamefont {Tka{\v{c}}ik}\ and\ \citenamefont
  {Walczak}(2011)}]{tkavcik2011information}%
  \BibitemOpen
  \bibfield  {author} {\bibinfo {author} {\bibfnamefont {G.}~\bibnamefont
  {Tka{\v{c}}ik}}\ and\ \bibinfo {author} {\bibfnamefont {A.~M.}\ \bibnamefont
  {Walczak}},\ }\bibfield  {title} {\bibinfo {title} {Information transmission
  in genetic regulatory networks: a review},\ }\href@noop {} {\bibfield
  {journal} {\bibinfo  {journal} {Journal of Physics: Condensed Matter}\
  }\textbf {\bibinfo {volume} {23}},\ \bibinfo {pages} {153102} (\bibinfo
  {year} {2011})}\BibitemShut {NoStop}%
\bibitem [{\citenamefont {Cheong}\ \emph {et~al.}(2011)\citenamefont {Cheong},
  \citenamefont {Rhee}, \citenamefont {Wang}, \citenamefont {Nemenman},\ and\
  \citenamefont {Levchenko}}]{cheong2011information}%
  \BibitemOpen
  \bibfield  {author} {\bibinfo {author} {\bibfnamefont {R.}~\bibnamefont
  {Cheong}}, \bibinfo {author} {\bibfnamefont {A.}~\bibnamefont {Rhee}},
  \bibinfo {author} {\bibfnamefont {C.~J.}\ \bibnamefont {Wang}}, \bibinfo
  {author} {\bibfnamefont {I.}~\bibnamefont {Nemenman}},\ and\ \bibinfo
  {author} {\bibfnamefont {A.}~\bibnamefont {Levchenko}},\ }\bibfield  {title}
  {\bibinfo {title} {Information transduction capacity of noisy biochemical
  signaling networks},\ }\href@noop {} {\bibfield  {journal} {\bibinfo
  {journal} {science}\ }\textbf {\bibinfo {volume} {334}},\ \bibinfo {pages}
  {354} (\bibinfo {year} {2011})}\BibitemShut {NoStop}%
\bibitem [{\citenamefont {Selimkhanov}\ \emph {et~al.}(2014)\citenamefont
  {Selimkhanov}, \citenamefont {Taylor}, \citenamefont {Yao}, \citenamefont
  {Pilko}, \citenamefont {Albeck}, \citenamefont {Hoffmann}, \citenamefont
  {Tsimring},\ and\ \citenamefont {Wollman}}]{selimkhanov2014accurate}%
  \BibitemOpen
  \bibfield  {author} {\bibinfo {author} {\bibfnamefont {J.}~\bibnamefont
  {Selimkhanov}}, \bibinfo {author} {\bibfnamefont {B.}~\bibnamefont {Taylor}},
  \bibinfo {author} {\bibfnamefont {J.}~\bibnamefont {Yao}}, \bibinfo {author}
  {\bibfnamefont {A.}~\bibnamefont {Pilko}}, \bibinfo {author} {\bibfnamefont
  {J.}~\bibnamefont {Albeck}}, \bibinfo {author} {\bibfnamefont
  {A.}~\bibnamefont {Hoffmann}}, \bibinfo {author} {\bibfnamefont
  {L.}~\bibnamefont {Tsimring}},\ and\ \bibinfo {author} {\bibfnamefont
  {R.}~\bibnamefont {Wollman}},\ }\bibfield  {title} {\bibinfo {title}
  {Accurate information transmission through dynamic biochemical signaling
  networks},\ }\href@noop {} {\bibfield  {journal} {\bibinfo  {journal}
  {Science}\ }\textbf {\bibinfo {volume} {346}},\ \bibinfo {pages} {1370}
  (\bibinfo {year} {2014})}\BibitemShut {NoStop}%
\bibitem [{\citenamefont {Tka{\v{c}}ik}\ and\ \citenamefont
  {Ten~Wolde}(2025)}]{tkavcik2025information}%
  \BibitemOpen
  \bibfield  {author} {\bibinfo {author} {\bibfnamefont {G.}~\bibnamefont
  {Tka{\v{c}}ik}}\ and\ \bibinfo {author} {\bibfnamefont {P.~R.}\ \bibnamefont
  {Ten~Wolde}},\ }\bibfield  {title} {\bibinfo {title} {Information processing
  in biochemical networks},\ }\href@noop {} {\bibfield  {journal} {\bibinfo
  {journal} {Annual review of biophysics}\ }\textbf {\bibinfo {volume} {54}},\
  \bibinfo {pages} {249} (\bibinfo {year} {2025})}\BibitemShut {NoStop}%
\bibitem [{\citenamefont {Southworth}\ \emph {et~al.}(2009)\citenamefont
  {Southworth}, \citenamefont {Owen},\ and\ \citenamefont
  {Kim}}]{southworth2009aging}%
  \BibitemOpen
  \bibfield  {author} {\bibinfo {author} {\bibfnamefont {L.~K.}\ \bibnamefont
  {Southworth}}, \bibinfo {author} {\bibfnamefont {A.~B.}\ \bibnamefont
  {Owen}},\ and\ \bibinfo {author} {\bibfnamefont {S.~K.}\ \bibnamefont
  {Kim}},\ }\bibfield  {title} {\bibinfo {title} {Aging mice show a decreasing
  correlation of gene expression within genetic modules},\ }\href@noop {}
  {\bibfield  {journal} {\bibinfo  {journal} {PLoS genetics}\ }\textbf
  {\bibinfo {volume} {5}},\ \bibinfo {pages} {e1000776} (\bibinfo {year}
  {2009})}\BibitemShut {NoStop}%
\bibitem [{\citenamefont {Clemens}\ \emph {et~al.}(2021)\citenamefont
  {Clemens}, \citenamefont {Sivakumar}, \citenamefont {Pius}, \citenamefont
  {Sahu}, \citenamefont {Shinde}, \citenamefont {Mamiya}, \citenamefont
  {Luketich}, \citenamefont {Cui}, \citenamefont {Dixit}, \citenamefont {Hoeck}
  \emph {et~al.}}]{clemens2021biphasic}%
  \BibitemOpen
  \bibfield  {author} {\bibinfo {author} {\bibfnamefont {Z.}~\bibnamefont
  {Clemens}}, \bibinfo {author} {\bibfnamefont {S.}~\bibnamefont {Sivakumar}},
  \bibinfo {author} {\bibfnamefont {A.}~\bibnamefont {Pius}}, \bibinfo {author}
  {\bibfnamefont {A.}~\bibnamefont {Sahu}}, \bibinfo {author} {\bibfnamefont
  {S.}~\bibnamefont {Shinde}}, \bibinfo {author} {\bibfnamefont
  {H.}~\bibnamefont {Mamiya}}, \bibinfo {author} {\bibfnamefont
  {N.}~\bibnamefont {Luketich}}, \bibinfo {author} {\bibfnamefont
  {J.}~\bibnamefont {Cui}}, \bibinfo {author} {\bibfnamefont {P.}~\bibnamefont
  {Dixit}}, \bibinfo {author} {\bibfnamefont {J.~D.}\ \bibnamefont {Hoeck}},
  \emph {et~al.},\ }\bibfield  {title} {\bibinfo {title} {The biphasic and
  age-dependent impact of klotho on hallmarks of aging and skeletal muscle
  function},\ }\href@noop {} {\bibfield  {journal} {\bibinfo  {journal}
  {Elife}\ }\textbf {\bibinfo {volume} {10}},\ \bibinfo {pages} {e61138}
  (\bibinfo {year} {2021})}\BibitemShut {NoStop}%
\bibitem [{\citenamefont {Leote}\ \emph {et~al.}(2024)\citenamefont {Leote},
  \citenamefont {Lopes},\ and\ \citenamefont {Beyer}}]{leote2024loss}%
  \BibitemOpen
  \bibfield  {author} {\bibinfo {author} {\bibfnamefont {A.~C.}\ \bibnamefont
  {Leote}}, \bibinfo {author} {\bibfnamefont {F.}~\bibnamefont {Lopes}},\ and\
  \bibinfo {author} {\bibfnamefont {A.}~\bibnamefont {Beyer}},\ }\bibfield
  {title} {\bibinfo {title} {Loss of coordination between basic cellular
  processes in human aging},\ }\href@noop {} {\bibfield  {journal} {\bibinfo
  {journal} {Nature Aging}\ }\textbf {\bibinfo {volume} {4}},\ \bibinfo {pages}
  {1432} (\bibinfo {year} {2024})}\BibitemShut {NoStop}%
\bibitem [{\citenamefont {Wang}\ \emph {et~al.}(2025)\citenamefont {Wang},
  \citenamefont {Xiao}, \citenamefont {Zhao}, \citenamefont {Su}, \citenamefont
  {Xia}, \citenamefont {Yang}, \citenamefont {Ma},\ and\ \citenamefont
  {Kong}}]{wang2025personalized}%
  \BibitemOpen
  \bibfield  {author} {\bibinfo {author} {\bibfnamefont {H.-T.}\ \bibnamefont
  {Wang}}, \bibinfo {author} {\bibfnamefont {F.-H.}\ \bibnamefont {Xiao}},
  \bibinfo {author} {\bibfnamefont {L.}~\bibnamefont {Zhao}}, \bibinfo {author}
  {\bibfnamefont {Q.}~\bibnamefont {Su}}, \bibinfo {author} {\bibfnamefont
  {T.-R.}\ \bibnamefont {Xia}}, \bibinfo {author} {\bibfnamefont {L.-Q.}\
  \bibnamefont {Yang}}, \bibinfo {author} {\bibfnamefont {S.-Y.}\ \bibnamefont
  {Ma}},\ and\ \bibinfo {author} {\bibfnamefont {Q.-P.}\ \bibnamefont {Kong}},\
  }\bibfield  {title} {\bibinfo {title} {Personalized transcriptional network
  analysis links age-related loss of gene coordination to individual biological
  aging},\ }\href@noop {} {\bibfield  {journal} {\bibinfo  {journal} {Genome
  Medicine}\ }\textbf {\bibinfo {volume} {17}},\ \bibinfo {pages} {1} (\bibinfo
  {year} {2025})}\BibitemShut {NoStop}%
\bibitem [{\citenamefont {Sivakumar}\ \emph {et~al.}(2025)\citenamefont
  {Sivakumar}, \citenamefont {LeFebre}, \citenamefont {Menichetti},
  \citenamefont {Iijima}, \citenamefont {Mugler},\ and\ \citenamefont
  {Ambrosio}}]{sivakumar2025novel}%
  \BibitemOpen
  \bibfield  {author} {\bibinfo {author} {\bibfnamefont {S.}~\bibnamefont
  {Sivakumar}}, \bibinfo {author} {\bibfnamefont {R.~W.}\ \bibnamefont
  {LeFebre}}, \bibinfo {author} {\bibfnamefont {G.}~\bibnamefont {Menichetti}},
  \bibinfo {author} {\bibfnamefont {H.}~\bibnamefont {Iijima}}, \bibinfo
  {author} {\bibfnamefont {A.}~\bibnamefont {Mugler}},\ and\ \bibinfo {author}
  {\bibfnamefont {F.}~\bibnamefont {Ambrosio}},\ }\bibfield  {title} {\bibinfo
  {title} {A novel information-theoretic approach reveals loss of effective
  biomolecular communication in aging muscle cells},\ }\href@noop {} {\bibfield
   {journal} {\bibinfo  {journal} {The Journals of Gerontology, Series A:
  Biological Sciences and Medical Sciences}\ }\textbf {\bibinfo {volume}
  {80}},\ \bibinfo {pages} {glaf195} (\bibinfo {year} {2025})}\BibitemShut
  {NoStop}%
\bibitem [{\citenamefont {Emison}\ \emph {et~al.}(2025)\citenamefont {Emison},
  \citenamefont {Lynn}, \citenamefont {Mugler}, \citenamefont {Ambrosio},\ and\
  \citenamefont {Dixit}}]{emison2025decay}%
  \BibitemOpen
  \bibfield  {author} {\bibinfo {author} {\bibfnamefont {B.}~\bibnamefont
  {Emison}}, \bibinfo {author} {\bibfnamefont {C.~W.}\ \bibnamefont {Lynn}},
  \bibinfo {author} {\bibfnamefont {A.}~\bibnamefont {Mugler}}, \bibinfo
  {author} {\bibfnamefont {F.}~\bibnamefont {Ambrosio}},\ and\ \bibinfo
  {author} {\bibfnamefont {P.~D.}\ \bibnamefont {Dixit}},\ }\bibfield  {title}
  {\bibinfo {title} {Decay in transcriptional information flow is a hallmark of
  cellular aging},\ }\href@noop {} {\bibfield  {journal} {\bibinfo  {journal}
  {bioRxiv}\ ,\ \bibinfo {pages} {2025}} (\bibinfo {year} {2025})}\BibitemShut
  {NoStop}%
\bibitem [{\citenamefont {Han}\ \emph {et~al.}(2018)\citenamefont {Han},
  \citenamefont {Cho}, \citenamefont {Lee}, \citenamefont {Yun}, \citenamefont
  {Kim}, \citenamefont {Bae}, \citenamefont {Yang}, \citenamefont {Kim},
  \citenamefont {Lee}, \citenamefont {Kim} \emph {et~al.}}]{han2018trrust}%
  \BibitemOpen
  \bibfield  {author} {\bibinfo {author} {\bibfnamefont {H.}~\bibnamefont
  {Han}}, \bibinfo {author} {\bibfnamefont {J.-W.}\ \bibnamefont {Cho}},
  \bibinfo {author} {\bibfnamefont {S.}~\bibnamefont {Lee}}, \bibinfo {author}
  {\bibfnamefont {A.}~\bibnamefont {Yun}}, \bibinfo {author} {\bibfnamefont
  {H.}~\bibnamefont {Kim}}, \bibinfo {author} {\bibfnamefont {D.}~\bibnamefont
  {Bae}}, \bibinfo {author} {\bibfnamefont {S.}~\bibnamefont {Yang}}, \bibinfo
  {author} {\bibfnamefont {C.~Y.}\ \bibnamefont {Kim}}, \bibinfo {author}
  {\bibfnamefont {M.}~\bibnamefont {Lee}}, \bibinfo {author} {\bibfnamefont
  {E.}~\bibnamefont {Kim}}, \emph {et~al.},\ }\bibfield  {title} {\bibinfo
  {title} {Trrust v2: an expanded reference database of human and mouse
  transcriptional regulatory interactions},\ }\href@noop {} {\bibfield
  {journal} {\bibinfo  {journal} {Nucleic acids research}\ }\textbf {\bibinfo
  {volume} {46}},\ \bibinfo {pages} {D380} (\bibinfo {year}
  {2018})}\BibitemShut {NoStop}%
\bibitem [{\citenamefont {Kang}\ \emph {et~al.}(2022)\citenamefont {Kang},
  \citenamefont {Tang}, \citenamefont {He}, \citenamefont {Li}, \citenamefont
  {Yang}, \citenamefont {Yu}, \citenamefont {Wang}, \citenamefont {Zhang},
  \citenamefont {Lin}, \citenamefont {Cui} \emph {et~al.}}]{kang2022rnainter}%
  \BibitemOpen
  \bibfield  {author} {\bibinfo {author} {\bibfnamefont {J.}~\bibnamefont
  {Kang}}, \bibinfo {author} {\bibfnamefont {Q.}~\bibnamefont {Tang}}, \bibinfo
  {author} {\bibfnamefont {J.}~\bibnamefont {He}}, \bibinfo {author}
  {\bibfnamefont {L.}~\bibnamefont {Li}}, \bibinfo {author} {\bibfnamefont
  {N.}~\bibnamefont {Yang}}, \bibinfo {author} {\bibfnamefont {S.}~\bibnamefont
  {Yu}}, \bibinfo {author} {\bibfnamefont {M.}~\bibnamefont {Wang}}, \bibinfo
  {author} {\bibfnamefont {Y.}~\bibnamefont {Zhang}}, \bibinfo {author}
  {\bibfnamefont {J.}~\bibnamefont {Lin}}, \bibinfo {author} {\bibfnamefont
  {T.}~\bibnamefont {Cui}}, \emph {et~al.},\ }\bibfield  {title} {\bibinfo
  {title} {Rnainter v4. 0: Rna interactome repository with redefined confidence
  scoring system and improved accessibility},\ }\href@noop {} {\bibfield
  {journal} {\bibinfo  {journal} {Nucleic acids research}\ }\textbf {\bibinfo
  {volume} {50}},\ \bibinfo {pages} {D326} (\bibinfo {year}
  {2022})}\BibitemShut {NoStop}%
\bibitem [{\citenamefont {Li}\ \emph {et~al.}(2025)\citenamefont {Li},
  \citenamefont {Wang}, \citenamefont {Wang}, \citenamefont {Li},\ and\
  \citenamefont {Liu}}]{li2025regnetwork}%
  \BibitemOpen
  \bibfield  {author} {\bibinfo {author} {\bibfnamefont {B.}~\bibnamefont
  {Li}}, \bibinfo {author} {\bibfnamefont {C.}~\bibnamefont {Wang}}, \bibinfo
  {author} {\bibfnamefont {Y.}~\bibnamefont {Wang}}, \bibinfo {author}
  {\bibfnamefont {P.}~\bibnamefont {Li}},\ and\ \bibinfo {author}
  {\bibfnamefont {Z.-P.}\ \bibnamefont {Liu}},\ }\bibfield  {title} {\bibinfo
  {title} {Regnetwork 2025: an integrative data repository for gene regulatory
  networks in human and mouse},\ }\href@noop {} {\bibfield  {journal} {\bibinfo
   {journal} {Nucleic Acids Research}\ ,\ \bibinfo {pages} {gkaf779}} (\bibinfo
  {year} {2025})}\BibitemShut {NoStop}%
\bibitem [{\citenamefont {Kimmel}\ \emph {et~al.}(2019)\citenamefont {Kimmel},
  \citenamefont {Penland}, \citenamefont {Rubinstein}, \citenamefont
  {Hendrickson}, \citenamefont {Kelley},\ and\ \citenamefont
  {Rosenthal}}]{kimmel2019murine}%
  \BibitemOpen
  \bibfield  {author} {\bibinfo {author} {\bibfnamefont {J.~C.}\ \bibnamefont
  {Kimmel}}, \bibinfo {author} {\bibfnamefont {L.}~\bibnamefont {Penland}},
  \bibinfo {author} {\bibfnamefont {N.~D.}\ \bibnamefont {Rubinstein}},
  \bibinfo {author} {\bibfnamefont {D.~G.}\ \bibnamefont {Hendrickson}},
  \bibinfo {author} {\bibfnamefont {D.~R.}\ \bibnamefont {Kelley}},\ and\
  \bibinfo {author} {\bibfnamefont {A.~Z.}\ \bibnamefont {Rosenthal}},\
  }\bibfield  {title} {\bibinfo {title} {Murine single-cell rna-seq reveals
  cell-identity-and tissue-specific trajectories of aging},\ }\href@noop {}
  {\bibfield  {journal} {\bibinfo  {journal} {Genome research}\ }\textbf
  {\bibinfo {volume} {29}},\ \bibinfo {pages} {2088} (\bibinfo {year}
  {2019})}\BibitemShut {NoStop}%
\bibitem [{\citenamefont {{The Tabula Muris
  Consortium}}(2020)}]{tabula2020single}%
  \BibitemOpen
  \bibfield  {author} {\bibinfo {author} {\bibnamefont {{The Tabula Muris
  Consortium}}},\ }\bibfield  {title} {\bibinfo {title} {A single-cell
  transcriptomic atlas characterizes ageing tissues in the mouse},\ }\href@noop
  {} {\bibfield  {journal} {\bibinfo  {journal} {Nature}\ }\textbf {\bibinfo
  {volume} {583}},\ \bibinfo {pages} {590} (\bibinfo {year}
  {2020})}\BibitemShut {NoStop}%
\bibitem [{\citenamefont {Krishnarajah}\ \emph {et~al.}(2022)\citenamefont
  {Krishnarajah}, \citenamefont {Ingelfinger}, \citenamefont {Friebel},
  \citenamefont {Cansever}, \citenamefont {Amorim}, \citenamefont {Andreadou},
  \citenamefont {Bamert}, \citenamefont {Litscher}, \citenamefont {Lutz},
  \citenamefont {Mayoux} \emph {et~al.}}]{krishnarajah2022single}%
  \BibitemOpen
  \bibfield  {author} {\bibinfo {author} {\bibfnamefont {S.}~\bibnamefont
  {Krishnarajah}}, \bibinfo {author} {\bibfnamefont {F.}~\bibnamefont
  {Ingelfinger}}, \bibinfo {author} {\bibfnamefont {E.}~\bibnamefont
  {Friebel}}, \bibinfo {author} {\bibfnamefont {D.}~\bibnamefont {Cansever}},
  \bibinfo {author} {\bibfnamefont {A.}~\bibnamefont {Amorim}}, \bibinfo
  {author} {\bibfnamefont {M.}~\bibnamefont {Andreadou}}, \bibinfo {author}
  {\bibfnamefont {D.}~\bibnamefont {Bamert}}, \bibinfo {author} {\bibfnamefont
  {G.}~\bibnamefont {Litscher}}, \bibinfo {author} {\bibfnamefont
  {M.}~\bibnamefont {Lutz}}, \bibinfo {author} {\bibfnamefont {M.}~\bibnamefont
  {Mayoux}}, \emph {et~al.},\ }\bibfield  {title} {\bibinfo {title}
  {Single-cell profiling of immune system alterations in lymphoid, barrier and
  solid tissues in aged mice},\ }\href@noop {} {\bibfield  {journal} {\bibinfo
  {journal} {Nature Aging}\ }\textbf {\bibinfo {volume} {2}},\ \bibinfo {pages}
  {74} (\bibinfo {year} {2022})}\BibitemShut {NoStop}%
\bibitem [{\citenamefont {Lagger}\ \emph {et~al.}(2023)\citenamefont {Lagger},
  \citenamefont {Ursu}, \citenamefont {Equey}, \citenamefont {Avelar},
  \citenamefont {Pisco}, \citenamefont {Tacutu},\ and\ \citenamefont
  {de~Magalh{\~a}es}}]{lagger2023scdiffcom}%
  \BibitemOpen
  \bibfield  {author} {\bibinfo {author} {\bibfnamefont {C.}~\bibnamefont
  {Lagger}}, \bibinfo {author} {\bibfnamefont {E.}~\bibnamefont {Ursu}},
  \bibinfo {author} {\bibfnamefont {A.}~\bibnamefont {Equey}}, \bibinfo
  {author} {\bibfnamefont {R.~A.}\ \bibnamefont {Avelar}}, \bibinfo {author}
  {\bibfnamefont {A.~O.}\ \bibnamefont {Pisco}}, \bibinfo {author}
  {\bibfnamefont {R.}~\bibnamefont {Tacutu}},\ and\ \bibinfo {author}
  {\bibfnamefont {J.~P.}\ \bibnamefont {de~Magalh{\~a}es}},\ }\bibfield
  {title} {\bibinfo {title} {scdiffcom: a tool for differential analysis of
  cell--cell interactions provides a mouse atlas of aging changes in
  intercellular communication},\ }\href@noop {} {\bibfield  {journal} {\bibinfo
   {journal} {Nature Aging}\ }\textbf {\bibinfo {volume} {3}},\ \bibinfo
  {pages} {1446} (\bibinfo {year} {2023})}\BibitemShut {NoStop}%
\bibitem [{\citenamefont {Xu}\ \emph {et~al.}(2025)\citenamefont {Xu},
  \citenamefont {Li}, \citenamefont {Hu}, \citenamefont {Luo}, \citenamefont
  {Gao}, \citenamefont {Li}, \citenamefont {Li},\ and\ \citenamefont
  {Zhang}}]{xu2025developing}%
  \BibitemOpen
  \bibfield  {author} {\bibinfo {author} {\bibfnamefont {Y.}~\bibnamefont
  {Xu}}, \bibinfo {author} {\bibfnamefont {M.}~\bibnamefont {Li}}, \bibinfo
  {author} {\bibfnamefont {C.}~\bibnamefont {Hu}}, \bibinfo {author}
  {\bibfnamefont {Y.}~\bibnamefont {Luo}}, \bibinfo {author} {\bibfnamefont
  {X.}~\bibnamefont {Gao}}, \bibinfo {author} {\bibfnamefont {X.}~\bibnamefont
  {Li}}, \bibinfo {author} {\bibfnamefont {X.}~\bibnamefont {Li}},\ and\
  \bibinfo {author} {\bibfnamefont {Y.}~\bibnamefont {Zhang}},\ }\bibfield
  {title} {\bibinfo {title} {Developing a novel aging assessment model to
  uncover heterogeneity in organ aging and screening of aging-related drugs},\
  }\href@noop {} {\bibfield  {journal} {\bibinfo  {journal} {Genome Medicine}\
  }\textbf {\bibinfo {volume} {17}},\ \bibinfo {pages} {83} (\bibinfo {year}
  {2025})}\BibitemShut {NoStop}%
\bibitem [{cod()}]{code}%
  \BibitemOpen
  \href@noop {} {\bibinfo {title} {Code and data available at
  \url{https://doi.org/10.5281/zenodo.18157674}}}\BibitemShut {NoStop}%
\bibitem [{\citenamefont {L{\"a}hnemann}\ \emph {et~al.}(2020)\citenamefont
  {L{\"a}hnemann}, \citenamefont {K{\"o}ster}, \citenamefont {Szczurek},
  \citenamefont {McCarthy}, \citenamefont {Hicks}, \citenamefont {Robinson},
  \citenamefont {Vallejos}, \citenamefont {Campbell}, \citenamefont
  {Beerenwinkel}, \citenamefont {Mahfouz} \emph
  {et~al.}}]{lahnemann2020eleven}%
  \BibitemOpen
  \bibfield  {author} {\bibinfo {author} {\bibfnamefont {D.}~\bibnamefont
  {L{\"a}hnemann}}, \bibinfo {author} {\bibfnamefont {J.}~\bibnamefont
  {K{\"o}ster}}, \bibinfo {author} {\bibfnamefont {E.}~\bibnamefont
  {Szczurek}}, \bibinfo {author} {\bibfnamefont {D.~J.}\ \bibnamefont
  {McCarthy}}, \bibinfo {author} {\bibfnamefont {S.~C.}\ \bibnamefont {Hicks}},
  \bibinfo {author} {\bibfnamefont {M.~D.}\ \bibnamefont {Robinson}}, \bibinfo
  {author} {\bibfnamefont {C.~A.}\ \bibnamefont {Vallejos}}, \bibinfo {author}
  {\bibfnamefont {K.~R.}\ \bibnamefont {Campbell}}, \bibinfo {author}
  {\bibfnamefont {N.}~\bibnamefont {Beerenwinkel}}, \bibinfo {author}
  {\bibfnamefont {A.}~\bibnamefont {Mahfouz}}, \emph {et~al.},\ }\bibfield
  {title} {\bibinfo {title} {Eleven grand challenges in single-cell data
  science},\ }\href@noop {} {\bibfield  {journal} {\bibinfo  {journal} {Genome
  biology}\ }\textbf {\bibinfo {volume} {21}},\ \bibinfo {pages} {31} (\bibinfo
  {year} {2020})}\BibitemShut {NoStop}%
\bibitem [{\citenamefont {Qiu}(2020)}]{qiu2020embracing}%
  \BibitemOpen
  \bibfield  {author} {\bibinfo {author} {\bibfnamefont {P.}~\bibnamefont
  {Qiu}},\ }\bibfield  {title} {\bibinfo {title} {Embracing the dropouts in
  single-cell rna-seq analysis},\ }\href@noop {} {\bibfield  {journal}
  {\bibinfo  {journal} {Nature communications}\ }\textbf {\bibinfo {volume}
  {11}},\ \bibinfo {pages} {1169} (\bibinfo {year} {2020})}\BibitemShut
  {NoStop}%
\bibitem [{\citenamefont {Bouland}\ \emph {et~al.}(2023)\citenamefont
  {Bouland}, \citenamefont {Mahfouz},\ and\ \citenamefont
  {Reinders}}]{bouland2023consequences}%
  \BibitemOpen
  \bibfield  {author} {\bibinfo {author} {\bibfnamefont {G.~A.}\ \bibnamefont
  {Bouland}}, \bibinfo {author} {\bibfnamefont {A.}~\bibnamefont {Mahfouz}},\
  and\ \bibinfo {author} {\bibfnamefont {M.~J.}\ \bibnamefont {Reinders}},\
  }\bibfield  {title} {\bibinfo {title} {Consequences and opportunities arising
  due to sparser single-cell rna-seq datasets},\ }\href@noop {} {\bibfield
  {journal} {\bibinfo  {journal} {Genome biology}\ }\textbf {\bibinfo {volume}
  {24}},\ \bibinfo {pages} {86} (\bibinfo {year} {2023})}\BibitemShut {NoStop}%
\bibitem [{\citenamefont {Sarra}\ \emph {et~al.}(2025)\citenamefont {Sarra},
  \citenamefont {Sarra}, \citenamefont {Di~Carlo}, \citenamefont {GrandPre},
  \citenamefont {Zhang}, \citenamefont {Callan~Jr},\ and\ \citenamefont
  {Bialek}}]{sarra2025maximum}%
  \BibitemOpen
  \bibfield  {author} {\bibinfo {author} {\bibfnamefont {C.}~\bibnamefont
  {Sarra}}, \bibinfo {author} {\bibfnamefont {L.}~\bibnamefont {Sarra}},
  \bibinfo {author} {\bibfnamefont {L.}~\bibnamefont {Di~Carlo}}, \bibinfo
  {author} {\bibfnamefont {T.}~\bibnamefont {GrandPre}}, \bibinfo {author}
  {\bibfnamefont {Y.}~\bibnamefont {Zhang}}, \bibinfo {author} {\bibfnamefont
  {C.~G.}\ \bibnamefont {Callan~Jr}},\ and\ \bibinfo {author} {\bibfnamefont
  {W.}~\bibnamefont {Bialek}},\ }\bibfield  {title} {\bibinfo {title} {Maximum
  entropy models for patterns of gene expression},\ }\href@noop {} {\bibfield
  {journal} {\bibinfo  {journal} {Physical Review E}\ }\textbf {\bibinfo
  {volume} {112}},\ \bibinfo {pages} {014408} (\bibinfo {year}
  {2025})}\BibitemShut {NoStop}%
\bibitem [{\citenamefont {Shannon}(1948)}]{shannon1948mathematical}%
  \BibitemOpen
  \bibfield  {author} {\bibinfo {author} {\bibfnamefont {C.~E.}\ \bibnamefont
  {Shannon}},\ }\bibfield  {title} {\bibinfo {title} {A mathematical theory of
  communication},\ }\href@noop {} {\bibfield  {journal} {\bibinfo  {journal}
  {The Bell system technical journal}\ }\textbf {\bibinfo {volume} {27}},\
  \bibinfo {pages} {379} (\bibinfo {year} {1948})}\BibitemShut {NoStop}%
\bibitem [{\citenamefont {Cover}(1999)}]{cover1999elements}%
  \BibitemOpen
  \bibfield  {author} {\bibinfo {author} {\bibfnamefont {T.~M.}\ \bibnamefont
  {Cover}},\ }\href@noop {} {\emph {\bibinfo {title} {Elements of information
  theory}}}\ (\bibinfo  {publisher} {John Wiley \& Sons},\ \bibinfo {year}
  {1999})\BibitemShut {NoStop}%
\bibitem [{\citenamefont {Cai}\ \emph {et~al.}(2021)\citenamefont {Cai},
  \citenamefont {Liu}, \citenamefont {Song}, \citenamefont {Wang},
  \citenamefont {Xu}, \citenamefont {He}, \citenamefont {Zhang}, \citenamefont
  {Zhang}, \citenamefont {Cheng}, \citenamefont {Jin} \emph
  {et~al.}}]{cai2021deficiency}%
  \BibitemOpen
  \bibfield  {author} {\bibinfo {author} {\bibfnamefont {Y.}~\bibnamefont
  {Cai}}, \bibinfo {author} {\bibfnamefont {H.}~\bibnamefont {Liu}}, \bibinfo
  {author} {\bibfnamefont {E.}~\bibnamefont {Song}}, \bibinfo {author}
  {\bibfnamefont {L.}~\bibnamefont {Wang}}, \bibinfo {author} {\bibfnamefont
  {J.}~\bibnamefont {Xu}}, \bibinfo {author} {\bibfnamefont {Y.}~\bibnamefont
  {He}}, \bibinfo {author} {\bibfnamefont {D.}~\bibnamefont {Zhang}}, \bibinfo
  {author} {\bibfnamefont {L.}~\bibnamefont {Zhang}}, \bibinfo {author}
  {\bibfnamefont {K.~K.-y.}\ \bibnamefont {Cheng}}, \bibinfo {author}
  {\bibfnamefont {L.}~\bibnamefont {Jin}}, \emph {et~al.},\ }\bibfield  {title}
  {\bibinfo {title} {Deficiency of telomere-associated repressor activator
  protein 1 precipitates cardiac aging in mice via p53/ppar$\alpha$
  signaling},\ }\href@noop {} {\bibfield  {journal} {\bibinfo  {journal}
  {Theranostics}\ }\textbf {\bibinfo {volume} {11}},\ \bibinfo {pages} {4710}
  (\bibinfo {year} {2021})}\BibitemShut {NoStop}%
\bibitem [{\citenamefont {Kan}\ \emph {et~al.}(2021)\citenamefont {Kan},
  \citenamefont {Hu}, \citenamefont {Ge}, \citenamefont {Zhang}, \citenamefont
  {Lu}, \citenamefont {Zhao}, \citenamefont {Zhang},\ and\ \citenamefont
  {Liu}}]{kan2021declined}%
  \BibitemOpen
  \bibfield  {author} {\bibinfo {author} {\bibfnamefont {J.}~\bibnamefont
  {Kan}}, \bibinfo {author} {\bibfnamefont {Y.}~\bibnamefont {Hu}}, \bibinfo
  {author} {\bibfnamefont {Y.}~\bibnamefont {Ge}}, \bibinfo {author}
  {\bibfnamefont {W.}~\bibnamefont {Zhang}}, \bibinfo {author} {\bibfnamefont
  {S.}~\bibnamefont {Lu}}, \bibinfo {author} {\bibfnamefont {C.}~\bibnamefont
  {Zhao}}, \bibinfo {author} {\bibfnamefont {R.}~\bibnamefont {Zhang}},\ and\
  \bibinfo {author} {\bibfnamefont {Y.}~\bibnamefont {Liu}},\ }\bibfield
  {title} {\bibinfo {title} {Declined expressions of vast mitochondria-related
  genes represented by cycs and transcription factor esrra in skeletal muscle
  aging},\ }\href@noop {} {\bibfield  {journal} {\bibinfo  {journal}
  {Bioengineered}\ }\textbf {\bibinfo {volume} {12}},\ \bibinfo {pages} {3485}
  (\bibinfo {year} {2021})}\BibitemShut {NoStop}%
\bibitem [{\citenamefont {Qian}\ \emph {et~al.}(2024)\citenamefont {Qian},
  \citenamefont {Zhu}, \citenamefont {Deng}, \citenamefont {Liang},
  \citenamefont {Chen}, \citenamefont {Chen}, \citenamefont {Wang},
  \citenamefont {Liu}, \citenamefont {Tian},\ and\ \citenamefont
  {Yang}}]{qian2024peroxisome}%
  \BibitemOpen
  \bibfield  {author} {\bibinfo {author} {\bibfnamefont {L.}~\bibnamefont
  {Qian}}, \bibinfo {author} {\bibfnamefont {Y.}~\bibnamefont {Zhu}}, \bibinfo
  {author} {\bibfnamefont {C.}~\bibnamefont {Deng}}, \bibinfo {author}
  {\bibfnamefont {Z.}~\bibnamefont {Liang}}, \bibinfo {author} {\bibfnamefont
  {J.}~\bibnamefont {Chen}}, \bibinfo {author} {\bibfnamefont {Y.}~\bibnamefont
  {Chen}}, \bibinfo {author} {\bibfnamefont {X.}~\bibnamefont {Wang}}, \bibinfo
  {author} {\bibfnamefont {Y.}~\bibnamefont {Liu}}, \bibinfo {author}
  {\bibfnamefont {Y.}~\bibnamefont {Tian}},\ and\ \bibinfo {author}
  {\bibfnamefont {Y.}~\bibnamefont {Yang}},\ }\bibfield  {title} {\bibinfo
  {title} {Peroxisome proliferator-activated receptor gamma coactivator-1
  (pgc-1) family in physiological and pathophysiological process and
  diseases},\ }\href@noop {} {\bibfield  {journal} {\bibinfo  {journal} {Signal
  transduction and targeted therapy}\ }\textbf {\bibinfo {volume} {9}},\
  \bibinfo {pages} {50} (\bibinfo {year} {2024})}\BibitemShut {NoStop}%
\end{thebibliography}
%apsrev4-2.bst 2019-01-14 (MD) hand-edited version of apsrev4-1.bst
%Control: key (0)
%Control: author (8) initials jnrlst
%Control: editor formatted (1) identically to author
%Control: production of article title (0) allowed
%Control: page (0) single
%Control: year (1) truncated
%Control: production of eprint (0) enabled
%

\end{document}